\documentclass[prl,aps,twocolumn,superscriptaddress,showpacs]{revtex4}

\usepackage{graphicx,graphics,epsfig}
\usepackage{dcolumn}
\usepackage{bm}
\usepackage{amsmath}
\usepackage{verbatim}
\usepackage{color}
\usepackage[colorlinks=false]{hyperref} 
\usepackage{subfigure}
\usepackage{times,natbib}
\usepackage{amsmath,amsfonts,amssymb,graphics,graphicx,epsfig,color,times,natbib}

\newcommand{\tr}{\mbox{tr}}

\date{\today}

\begin{document}
\title{Hardy's Paradox for High-Dimensional Systems: Beyond Hardy's Limit}

\author{Jing-Ling Chen}
\email{chenjl@nankai.edu.cn}
 \affiliation{Theoretical Physics Division, Chern Institute of Mathematics, Nankai University,
 Tianjin 300071, People's Republic of China}
 \affiliation{Centre for Quantum Technologies, National University of Singapore,
 3 Science Drive 2, Singapore 117543}

\author{Ad\'an Cabello}
\email{adan@us.es}
\affiliation{Departamento de F\'{\i}sica Aplicada II, Universidad de Sevilla, E-41012 Sevilla, Spain}

\author{Zhen-Peng Xu}
 \affiliation{Theoretical Physics Division, Chern Institute of Mathematics, Nankai University,
 Tianjin 300071, People's Republic of China}

\author{Hong-Yi Su}
 \affiliation{Theoretical Physics Division, Chern Institute of Mathematics, Nankai University,
 Tianjin 300071, People's Republic of China}

\author{Chunfeng~Wu}
 \affiliation{Pillar of Engineering Product Development, Singapore University of Technology and Design,
  20 Dover Drive, Singapore 138682}

\author{L.~C.~Kwek}
\email{kwekleongchuan@nus.edu.sg}
 \affiliation{Centre for Quantum Technologies, National University of Singapore,
 3 Science Drive 2, Singapore 117543}
 \affiliation{National Institute of Education and Institute of Advanced Studies,
 Nanyang Technological University, 1 Nanyang Walk, Singapore 637616}


\begin{abstract}
 Hardy's proof is considered the simplest proof of nonlocality. Here we introduce an equally simple proof that (i) has Hardy's as a particular case, (ii) shows that the probability of nonlocal events grows with the dimension of the local systems, and (iii) is always equivalent to the violation of a tight Bell inequality.
\end{abstract}



\maketitle




{\em Introduction.---}Nonlocality, namely, the impossibility of describing correlations in terms of local hidden variables \cite{Bell64}, is a fundamental property of nature. Hardy's proof \cite{Hardy92,Hardy93}, in any of its forms \cite{Goldstein94,Mermin94a,Mermin94b,KH05}, provides a simple way to show that quantum correlations cannot be explained with local theories. Hardy's proof is usually considered ``the simplest form of Bell's theorem'' \cite{Mermin95}.

On the other hand, if one wants to study nonlocality in a systematic way, one must define the local polytope \cite{Pitowsky89} corresponding to any possible scenario (i.e., for any given number of parties, settings, and outcomes) and check whether quantum correlations violate the inequalities defining the facets of the corresponding local polytope. These inequalities are the so-called {\em tight} Bell inequalities. In this sense, Hardy's proof has another remarkable property: It is equivalent to a violation of a tight Bell inequality, the Clauser-Horne-Shimony-Holt (CHSH) inequality \cite{CHSH69}. This was observed in \cite{Mermin94a}.

Hardy's proof requires two observers, each with two measurements, each with two possible outcomes. The proof has been extended to the case of more than two measurements \cite{Hardy97,BBDH97}, and more than two outcomes \cite{KC05,SG11,RZS12}. However, none of these extensions is equivalent to the violation of a tight Bell inequality.

The aim of this Letter is to show that, if we remove the requirement that the measurements have two outcomes, then Hardy's proof can be formulated in a much powerful way. The new formulation shows that the maximum probability of nonlocal events, which has a limit of $0.09$ in Hardy's formulation and previously proposed extensions, actually grows with the number of possible outcomes, tending asymptotically to a limit that is more than four times higher than the original one. Moreover, for any given number of outcomes, the new formulation turns out to be equivalent to a violation of a tight Bell inequality, a feature that suggest that this formulation is more fundamental than any other one proposed previously. All this while preserving the simplicity of Hardy's original proof.




{\em A new formulation of Hardy's paradox.---}Let us consider two observers, Alice, who can measure either $A_1$ or $A_2$ on her subsystem, and Bob, who can measure $B_1$ or $B_2$ on his. Suppose that each of these measurements has $d$ outcomes that we will number as $0,1,2,\ldots,d-1$. Let us denote as $P(A_2 < B_1)$ the joint conditional probability that the result of $A_2$ is strictly smaller than the result of $B_1$, that is,
\begin{equation}
P(A_2 < B_1)=\sum_{m<n}P(A_2=m, B_1=n),
\end{equation}
with $m, n \in \{0, 1, \ldots, d-1\}$.
Explicitly, for $d=2$, $P(A_2 < B_1)=P(A_2=0,B_1=1)$; for $d=3$, $P(A_2 <
B_1)=P(A_2=0,B_1=1)+P(A_2=0,B_1=2)+P(A_2=1,B_1=2)$, etc.

Then, the proof follows from the fact that, according to quantum theory, there are two-qudit entangled states and local measurements satisfying, simultaneously, the following conditions:
\begin{subequations}
 \label{E1}
 \begin{align}
 &P(A_2 < B_1) = 0, \label{E1a}\\
 &P(B_1 < A_1) = 0, \label{E1b}\\
 &P(A_1 < B_2) = 0, \label{E1c}\\
 &P(A_2 < B_2) > 0. \label{E1d}
\end{align}
\end{subequations}
Therefore, if events $A_2<B_1$, $B_1<A_1$, and $A_1<B_2$ never happen, then, in any local theory, event $A_2
< B_2$ must never happen either. However, this is in contradiction with (\ref{E1d}).

If $d=2$, the proof is exactly Hardy's \cite{Hardy92,Hardy93}.


{\em Beyond Hardy's limit.---}Let us define,
\begin{equation}
 P_{\rm Hardy}=\max P(A_2 < B_2)
\end{equation}
satisfying conditions (\ref{E1a})--(\ref{E1c}). For $d=2$,
\begin{eqnarray}
\label{Hlimit}
 P^{(d=2)}_{\rm Hardy}=\frac{5\sqrt{5} - 11}{2}\approx 0.09,
\end{eqnarray}
and is achieved with two-qubit systems \cite{Hardy92,Hardy93}.

In previous extensions of Hardy's paradox to two-qudit systems \cite{KC05,SG11,RZS12}, (\ref{Hlimit}) is also the maximum probability of events that cannot be explained by local theories.

For example, the extension considered in Ref.~\cite{KC05} is based on the following four probabilities: $P(A_1 = 0, B_1 = 0) = 0$, $P(A_1 \neq 0, B_2 = 0) = 0$, $P(A_2 = 0, B_1 \neq 0) = 0$, and $P(A_2 = 0, B_2 = 0) = P_{\rm KC} > 0$.
Ref.~\cite{SG11} proves that, for two-qutrit systems, $\max P_{\rm KC}$ equals (\ref{Hlimit}), and conjectures that $\max P_{\rm KC}$ is always (\ref{Hlimit}) for arbitrary dimension. Ref.~\cite{RZS12} provides a proof of this conjecture.

Interestingly, in the proof presented in the previous section, $P_{\rm Hardy}$ equals Hardy's limit (\ref{Hlimit}) for $d=2$, but this is not longer true for higher dimensional systems.

To show this, we will
consider pure states
satisfying the three conditions (\ref{E1a})--(\ref{E1c}). An arbitrary two-qudit pure state can be
written as
\begin{eqnarray}
|\Psi\rangle = \sum\limits_{i = 0}^{d-1} \sum\limits_{j = 0}^{d-1}
h_{ij} |i\rangle_A|j\rangle_B,
\end{eqnarray}
where the basis states
$|i\rangle_A, |j\rangle_B \in \{|0\rangle, |1\rangle, \ldots, |d-1\rangle\}$,
and $h_{ij}$ are coefficients satisfying the normalization
condition $\sum_{ij} |h_{ij}|^2 = 1$.

The coefficients $h_{ij}$ completely determine
the state $|\Psi\rangle$. We can associate any two-qudit state $|\Psi\rangle$ with a coefficient-matrix
$H=(h_{ij})_{d \times d}$,
where $i, j=0, 1, \ldots, d-1$, and $h_{ij}$ is the $i$-th row and
the $j$-column element of the $d\times d$ matrix $H$.
The connection between
the coefficient-matrix $H$ and the two reduced density matrices of
$|\Psi\rangle\langle\Psi|$ is
\begin{subequations}
\begin{align}
\rho_A&=\tr_B(|\Psi\rangle\langle\Psi|)=HH^\dag, \\
\rho_B&=\tr_A(|\Psi\rangle\langle\Psi|)=H^T(H^T)^\dag,
\end{align}
\end{subequations}
where $T$ for matrix transpose and $H^\dag$ is the hermitian conjugate matrix of $H$.

The probability $P(A_i=m, B_j=n)$ can be calculated as
\begin{eqnarray}
\label{PRab}
 P(A_i=m, B_j=n)=\tr[(\hat{\Pi}_{A_i}^m\otimes
\hat{\Pi}_{B_j}^n)\rho],
\end{eqnarray}
where $\hat{\Pi}_{A_i}^m$ and $\hat{\Pi}_{B_j}^n$ are projectors, and $\rho=|\Psi\rangle\langle\Psi|$.
Explicitly, the projectors are given by
\begin{subequations}
 \label{PR1}
 \begin{align}
 &\hat{\Pi}_{A_1}^m=\mathcal {U}_1 \;|m\rangle\langle m| \;\mathcal {U}_1^\dag, \label{PR1a}\\
 &\hat{\Pi}_{B_1}^n=\mathcal {V}_1 \;|n\rangle\langle n| \;\mathcal {V}_1^\dag, \label{PR1b}\\
 &\hat{\Pi}_{A_2}^m=\mathcal {U}_2 \;|m\rangle\langle m| \;\mathcal {U}_2^\dag, \label{PR1c}\\
 &\hat{\Pi}_{B_2}^n=\mathcal {V}_2 \;|n\rangle\langle n| \;\mathcal {V}_2^\dag, \label{PR1d}
\end{align}
\end{subequations}
where
$\mathcal {U}_1$, $\mathcal {V}_1$, $\mathcal {U}_2$, and
$\mathcal {V}_2$
are, in general, $SU(d)$ unitary matrices.

To calculate $P^{\rm opt}_{\rm Hardy}$, it is sufficient to choose
the settings $A_1$ and $B_1$ as the standard bases, i.e., taking
$\mathcal {U}_1=\mathcal {V}_1=\openone$, where $\openone$ is the
identity matrix. Evidently, the condition (\ref{E1b}) leads to $h_{ij}=0$, for $i>j$.
This implies that the matrix $H$ is an upper-triangular
matrix.

In Table \ref{Table1}, we list the optimal values of $P^{\rm
opt}_{\rm Hardy}$ for $d=2,\dots, 7$. The corresponding  optimal
Hardy states $H^{\rm opt}$ are explicitly given in the Appendix.

\begin{figure}
\includegraphics[width=80mm]{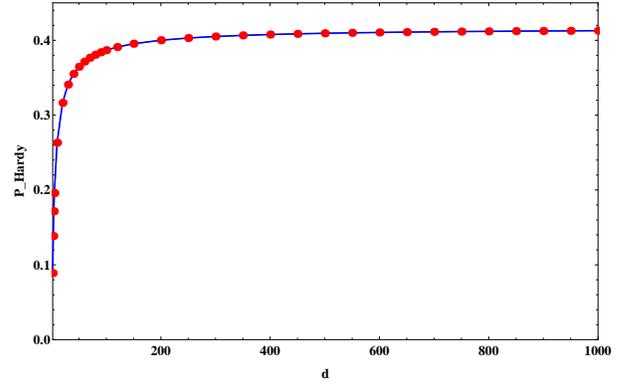}\\
 \caption{(Color online) $P^{\rm
app}_{\rm Hardy}$ from $d=2$ to $d=1000$.}\label{fig1}
\end{figure}

The calculations for $d>7$ are beyond our computers
capability. However, we observe that $H^{\rm opt}$, written in the
representation of $H$, have reflection symmetry with respect to the
anti-diagonal line, that is, $h_{ij}=h_{d-1-j, d-1-i}$. We use this
to calculate approximately the maximum probability for nonlocal
events $P^{\rm app}_{\rm Hardy}$, by using a special class of states
$H^{\rm app}$. The explicit form of states $H^{\rm app}$ is given in
the Appendix. This allows us to go beyond $d=7$ and
compute $P^{\rm app}_{\rm Hardy}$ from $d=2$ to $d=28000$. In Fig.
\ref{fig1}, we have plotted $P^{\rm app}_{\rm Hardy}$ from $d=2$ to
$d=1000$, showing that $P^{\rm app}_{\rm Hardy}$ increases with the dimension. Values for higher dimensions are given in the Appendix.

In Table \ref{Table1}, we also compare the $P_{\rm Hardy}$ for the
optimal states and the approximate optimal states. This allows us to
speculate that  the asymptotic limit may be a little higher than the
one showed in Fig.~\ref{fig1}. However, the limit $1/2$ can never be
surpassed since $P(A_2>B_2)$ is always bigger than $P(A_2<B_2)$ as observed in the numerical computations. At
this point, we do not know whether or not $1/2$ may be the
asymptotic limit.


\begin{table*}[t]
\centering
  \begin{tabular}{ c c c c c c c}
   $d$ & 2 & 3 & 4 & 5 & 6 & 7\\ \hline
   $P^{\rm opt}_{\rm Hardy}$ & 0.090170 & 0.141327 & 0.176512 & 0.203057 & 0.224221 & 0.241728\\
   $P^{\rm app}_{\rm Hardy}$ & 0.088889 & 0.138426 & 0.171533 & 0.195869 & 0.214825 & 0.230172 \\
   Error Rates & 0.014207 & 0.020527 & 0.020288 & 0.035399 & 0.0419051 & 0.047807\\
  \end{tabular}
  \caption{\label{Table1}$P^{\rm opt}_{\rm Hardy}$ and $P^{\rm app}_{\rm Hardy}$ for
$d=2,\ldots,7$.}
\end{table*}


{\em Degree of entanglement.---}Hardy's proof does not work for maximally entangled states. The same is true for the proof introduced here. However, in out proof, as $d$ increases, the degree of entanglement tends to 1. To show this, we use the generalized concurrence or degree of entanglement
\cite{AF01} for two-qudit systems given by
\begin{eqnarray}
\mathcal
{C}=\sqrt{\frac{d}{d-1}\biggr[1-\tr(\rho_A^2)\biggr]}=\sqrt{\frac{d}{d-1}\biggr[1-\tr(\rho_B^2)\biggr]}.
\end{eqnarray}

In Table \ref{Table2}, we have plotted $\mathcal{C}$ for the optimal Hardy's states and the approximate Hardy's states. From Table \ref{Table2}, we observe that, for $d=2$, the optimal Hardy's state
occurs at $\mathcal {C}^{\rm opt}\approx 0.763932$, and this value increases
to $\mathcal {C}^{\rm opt}\approx 0.827702$ when $d=5$. For a fixed
$d$, the corresponding $\mathcal {C}^{\rm app}$ is larger than that
of $\mathcal {C}^{\rm app}$, and it also increases with the
dimension $d$. For $d=800$, $\mathcal {C}^{\rm app}\approx
0.998062$, and tends to 1 as $d$ grows.


\begin{table*}[t]
\centering
  \begin{tabular}{c c c c c c c}
   $d$ & 2 & 3 & 4 & 5 & 6 & 7\\
   \hline
   Optimal States  & 0.763932 & 0.793888 & 0.813483 & 0.827702 & 0.838679 & 0.847510 \\
   Approximate States  & 0.825885 & 0.845942 &0.861735 & 0.874459& 0.884926 & 0.893695 \\
  \end{tabular}
\caption{\label{Table2}Entanglement degrees of the optimal states and the
approximate optimal states for $d=2, \ldots, 7$.}
\end{table*}


Finally, we can prove that the proof cannot work for two-qudit maximally entangled states,
\begin{eqnarray}
|\Psi\rangle_{\rm MES}=\frac{1}{\sqrt{d}}\sum_{j=0}^{d-1}|j\rangle_A |j\rangle_B.
\end{eqnarray}

\emph{Proof: }$\tr[(\hat{\Pi}_{A_1}^m\otimes \hat{\Pi}_{B_1}^n)
|\Psi\rangle\langle\Psi|]$ can be expressed as
\begin{equation}
\tr[ (|m\rangle\langle m| \otimes
|n\rangle\langle n| )(\mathcal {U}^\dag_1\otimes \mathcal
{V}^\dag_1) |\Psi\rangle\langle\Psi| (\mathcal {U}_1\otimes \mathcal
{V}_1)].
\end{equation}
We will use
\begin{eqnarray}
H_{\rm MES}\mapsto |\Psi\rangle_{\rm MES}, \;\; H_{1}\mapsto
(\mathcal {U}^\dag_1\otimes \mathcal {V}^\dag_1) |\Psi\rangle_{\rm
MES}.
\end{eqnarray}
Taking into account that: (i) given a pure state
$H\mapsto|\Psi\rangle_{AB}$ and a local action $U$ acting on Alice
(the first part) and $V$ acting on Bob (the second part), then
\begin{eqnarray}
H'\mapsto(U \otimes V) |\Psi\rangle_{AB} = U H V^T.
\end{eqnarray}
(ii) Eq. (\ref{E1b}) requires $H'$ to
be an upper-triangular matrix, and (iii) $H_{\rm
MES}=\frac{1}{\sqrt{d}}\openone$. Then, we have the solution
\begin{equation}
\mathcal {U}_1\mathcal {V}^T_1=\mathcal {D}_1,\label{D1}
\end{equation}
where $\mathcal {D}_1={\rm
diag}(e^{i\chi_0},e^{i\chi_1},\ldots,e^{i\chi_{d-1}})$. Similarly,
from (\ref{E1a}) and (\ref{E1c}), we have
\begin{equation}
\mathcal {U}_1\mathcal {V}^T_2=\mathcal {D}_2,\;\mathcal
{U}_2\mathcal {V}^T_1=\mathcal {D}_3,\label{D23}
\end{equation}
where $\mathcal {D}_2,\mathcal {D}_3$ are diagonal matrices similar
to $\mathcal {D}_1$. From (\ref{D1}) and (\ref{D23}) we have
\begin{equation}
\mathcal {U}_2\mathcal {V}^T_2=\mathcal {D}_3\mathcal
{D}^\dagger_1\mathcal {D}_2,
\end{equation}
which directly leads to $P(A_2<B_2)=0$ for $|\Psi\rangle_{\rm MES}$.
This ends the proof.


{\em Connection to tight Bell inequalities.---}As it can be easily seen, for any $d$, our proof can be transformed into the following Bell inequality:
\begin{equation}
\label{ZGineq}
\begin{split}
 &P(A_2 < B_1)+P(B_1 < A_1) \\
 &+ P(A_1 < B_2)-P(A_2 < B_2) \stackrel{\mbox{\tiny{ LHV}}}{\geq} 0,
\end{split}
\end{equation}
where LHV indicates that the bound is satisfied by local hidden variable theories.
The interesting point is that, for any $d$, inequality (\ref{ZGineq}) is equivalent to the Zohren and Gill's
version \cite{ZG08} of the Collins-Gisin-Linden-Massar-Popescu inequalities (the plural because there is a different inequality for each $d$) \cite{CGLMP02}, which are tight Bell inequalities for any $d$ \cite{Masanes03}. This feature distinguishes our proof from any previously proposed nonlocality proof having Hardy's as a particular case.


{\em Conclusions.---}Hardy's proof is considered the simplest proof of nonlocality. Here we have introduced an equally simple proof that reveals much more about nonlocality in the case that the local systems are qudits. When $d=2$, the proof is exactly Hardy's, but for $d>2$ the probability of nonlocal events grows with $d$, so, for high $d$, this probability is more than four times larger than in Hardy's and in previous extensions to two-qudit systems. Interestingly, we have showed that, for any $d$, our proof is always equivalent to the violation of a tight Bell inequality. This suggests that ours is the most natural and powerful generalization of Hardy's paradox when higher-dimensional systems are considered.


\begin{acknowledgments}
J.L.C. is supported by the National Basic Research Program (973
Program) of China under Grant No.\ 2012CB921900 and the NSF of China
(Grant Nos.\ 10975075 and 11175089). A.C. is supported by
Project No.\ FIS2011-29400 with FEDER funds (MINECO, Spain). This
work is also partly supported by the National Research Foundation
and the Ministry of Education, Singapore.
\end{acknowledgments}



\appendix


\section{Appendix A: Optimal Hardy states}


The optimal Hardy states $H_d$ for $d = 2,\ldots,7$ are
\begin{widetext}
\begin{subequations}
\begin{eqnarray}
H_2 = \left(
\begin{array}{cc}
 0.618034 & 0.485868 \\
 0 & 0.618034 \\
\end{array}
\right),
\end{eqnarray}
\begin{eqnarray}
H_3 = \left(
\begin{array}{ccc}
 0.498328 & 0.316483 & 0.329301 \\
 0 & 0.441108 & 0.316483 \\
 0 & 0 & 0.498328 \\
\end{array}
\right),
\end{eqnarray}
\begin{eqnarray}
H_4 = \left(
\begin{array}{cccc}
 0.429796 & 0.262169 & 0.224332 & 0.249934 \\
 0 & 0.376021 & 0.217224 & 0.224332 \\
 0 & 0 & 0.376021 & 0.262169 \\
 0 & 0 & 0 & 0.429796 \\
\end{array}
\right),
\end{eqnarray}
\begin{eqnarray}
H_5 = \left(
\begin{array}{ccccc}
 0.383613 & 0.230044 & 0.189636 & 0.175427 & 0.201533 \\
 0 & 0.334102 & 0.185035 & 0.157012 & 0.175427 \\
 0 & 0 & 0.33072 & 0.185035 & 0.189636 \\
 0 & 0 & 0 & 0.334102 & 0.230044 \\
 0 & 0 & 0 & 0 & 0.383613 \\
\end{array}
\right),
\end{eqnarray}
\begin{eqnarray}
H_6 = \left(
\begin{array}{cccccc}
 0.349686 & 0.207877 & 0.16845 & 0.150559 & 0.144455 & 0.16883 \\
 0 & 0.303795 & 0.165105 & 0.134967 & 0.125208 & 0.144455 \\
 0 & 0 & 0.29972 & 0.160666 & 0.134967 & 0.150559 \\
 0 & 0 & 0 & 0.29972 & 0.165105 & 0.16845 \\
 0 & 0 & 0 & 0 & 0.303795 & 0.207877 \\
 0 & 0 & 0 & 0 & 0 & 0.349686 \\
\end{array}
\right),
\end{eqnarray}
\begin{eqnarray}
H_7 = \left(
\begin{array}{ccccccc}
 0.323377 & 0.191279 & 0.153539 & 0.135037 & 0.12545 & 0.122887 & 0.145233 \\
 0 & 0.280442 & 0.150851 & 0.121193 & 0.108665 & 0.104707 & 0.122887 \\
 0 & 0 & 0.276282 & 0.145271 & 0.117498 & 0.108665 & 0.12545 \\
 0 & 0 & 0 & 0.275414 & 0.145271 & 0.121193 & 0.135037 \\
 0 & 0 & 0 & 0 & 0.276282 & 0.150851 & 0.153539 \\
 0 & 0 & 0 & 0 & 0 & 0.280442 & 0.191279 \\
 0 & 0 & 0 & 0 & 0 & 0 & 0.323377 \\
\end{array}
\right).
\end{eqnarray}
\end{subequations}
\end{widetext}



\section{Appendix B: Approximate optimal Hardy states}


The form of $H_d$ for $d = 2,\ldots,7$ suggests to define the approximate optimal Hardy states as follows:
\begin{eqnarray}
H_d^{\rm app}=\left(
\begin{array}{cccccc}
\alpha_1 & \alpha_2 & \alpha_3 & \cdots & \alpha_{d-1} & \alpha_d\\
        & \alpha_1 & \alpha_2 & \cdots  & \alpha_{d-2} & \alpha_{d-1}\\
        &          & \ddots   & \ddots  & \vdots       & \vdots\\
        &          &          & \alpha_1& \alpha_2     &\alpha_3\\
        &          &          &         & \alpha_1     &\alpha_2\\
        &          &          &         &              &\alpha_1\\
\end{array}
\right),
\end{eqnarray}
where
\begin{eqnarray}
\alpha_r=\frac{\beta_r}{\sqrt{d+1-r}}, \;\; r=1, 2, \ldots, d,
\end{eqnarray}
with $\beta_r>0$ satisfying the following relations:
\begin{subequations}
\begin{align}
&\beta_1:\beta_2:\beta_3:\cdots:\beta_d=1:\frac{1}{2}:\frac{1}{3}:\cdots:\frac{1}{d},\\
&\sum_{r=1}^d \beta_r^2=1.
\end{align}
\end{subequations}

In Table~\ref{to28000} we have listed $P^{\rm app}_{\rm Hardy}$ up to $d=28000$.


\begin{table}[t]
\centering
  \begin{tabular}{c c c c c c c c}
$d$ & $P^{\rm app}_{\rm Hardy}$& $d$ & $P^{\rm app}_{\rm Hardy}$ & $d$ & $P^{\rm app}_{\rm Hardy}$ & $d$ & $P^{\rm app}_{\rm Hardy}$ \\ \hline
 2  & 0.088889 &  300 & 0.405106 & 2000 & 0.414711 & 10000 & 0.416300 \\
 10 & 0.263168 &  400 & 0.407749 & 2200 & 0.414885 & 11000 & 0.416339 \\
 20 & 0.316491 &  500 & 0.409394 & 2400 & 0.415031 & 12000 & 0.416371 \\
 30 & 0.340836 &  600 & 0.410520 & 2600 & 0.415156 & 13000 & 0.416398 \\
 40 & 0.355158 &  700 & 0.411341 & 2800 & 0.415263 & 14000 & 0.416421 \\
 50 & 0.364700 &  800 & 0.411966 & 3000 & 0.415357 & 16000 & 0.416459 \\
 60 & 0.371554 &  900 & 0.412459 & 4000 & 0.415687 & 18000 & 0.416489 \\
 70 & 0.376736 & 1000 & 0.412857 & 5000 & 0.415889 & 20000 & 0.416513 \\
 80 & 0.380803 & 1200 & 0.413464 & 6000 & 0.416024 & 22000 & 0.416533 \\
 90 & 0.384085 & 1400 & 0.413903 & 6000 & 0.416024 & 24000 & 0.416549 \\
100 & 0.386793 & 1600 & 0.414230 & 8000 & 0.416196 & 26000 & 0.416563 \\
200 & 0.400116 & 1800 & 0.414499 & 9000 & 0.416254 & 28000 & 0.416575 \\
  \end{tabular}
\caption{$P^{\rm app}_{\rm Hardy}$ from $d=2$ to $d=28000$.} \label{to28000}
\end{table}



\begin{thebibliography}{99}

\bibitem{Bell64}
 J. S. Bell,
 Physics (Long Island City, N.Y.) \textbf{1}, 195
 (1964).

\bibitem{Hardy92}
 L. Hardy,
 \href{http://prl.aps.org/abstract/PRL/v68/i20/p2981_1}{Phys. Rev. Lett. \textbf{68}, 2981 (1992).}

\bibitem{Hardy93}
 L. Hardy,
 \href{http://dx.doi.org/10.1103/PhysRevLett.71.1665}{Phys. Rev. Lett. \textbf{71}, 1665 (1993).}


\bibitem{Goldstein94}
 S. Goldstein,
 \href{http://dx.doi.org/10.1103/PhysRevLett.72.1951}{Phys. Rev. Lett. \textbf{72}, 1951 (1994).}

\bibitem{Mermin94a}
 N. D. Mermin,
 \href{http://www.physicstoday.org/resource/1/phtoad/v47/i6/p9_s1?isAuthorized=no}{Phys. Today \textbf{47}(6), 9 (1994)};
 \href{http://www.physicstoday.org/resource/1/phtoad/v47/i11/p119_s2?isAuthorized=no}{Phys. Today \textbf{47}(11), 119 (1994)}.

\bibitem{Mermin94b}
 N. D. Mermin,
 \href{http://ajp.aapt.org/resource/1/ajpias/v62/i10/p880_s1?isAuthorized=no}{Am. J. Phys. \textbf{62}, 880 (1994).}

\bibitem{KH05}
 P. G. Kwiat and L. Hardy,
 \href{http://ajp.aapt.org/resource/1/ajpias/v68/i1/p33_s1?isAuthorized=no}{Am. J. Phys. \textbf{68}, 33 (2000).}

\bibitem{Mermin95}
 N. D. Mermin,
 in {\em Fundamental Problems in Quantum Theory},
 edited by D. M. Greenberger and A. Zeilinger,
 \href{http://onlinelibrary.wiley.com/doi/10.1111/j.1749-6632.1995.tb39001.x/abstract}{Ann. N. Y. Acad. Sci. \textbf{755}, 616 (1995).}


\bibitem{Pitowsky89}
 I. Pitowsky,
 \emph{Quantum Probability--Quantum Logic}
 (Springer, New York, 1989).


\bibitem{CHSH69}
 J. F. Clauser, M. A. Horne, A. Shimony, and R. A. Holt,
 \href{http://prl.aps.org/abstract/PRL/v23/i15/p880_1}{Phys. Rev. Lett. \textbf{23}, 880 (1969).}


\bibitem{Hardy97}
 L. Hardy,
 in {\em New Developments on Fundamental Problems in Quantum Physics},
 edited by M. Ferrero and A. van der Merwe
 (Kluwer, Dordrecht, Holland, 1997), p.~163.

\bibitem{BBDH97}
 D. Boschi, S. Branca, F. De Martini, and L. Hardy,
 \href{http://prl.aps.org/abstract/PRL/v79/i15/p2755_1}{Phys. Rev. Lett. \textbf{79}, 2755 (1997).}


\bibitem{KC05}
 S. Kunkri and S. K. Choudhary,
 \href{http://dx.doi.org/10.1103/PhysRevA.72.022348}{Phys. Rev. A \textbf{72}, 022348 (2005).}

\bibitem{SG11}
 K. P. Seshadreesan and S. Ghosh,
 \href{http://dx.doi.org/10.1088/1751-8113/44/31/315305}{J. Phys. A: Math. Theor. \textbf{44}, 315305 (2011).}

\bibitem{RZS12}
 R. Rabelo, L. Y. Zhi, and V. Scarani,
 \href{http://prl.aps.org/abstract/PRL/v109/i18/e180401}{Phys. Rev. Lett. \textbf{109}, 180401 (2012).}

\bibitem{AF01}
 S. Albeverio and S.-M. Fei,
 \href{http://iopscience.iop.org/1464-4266/3/4/305/}{J. Opt. B: Quant. Semiclass. Opt. \textbf{3}, 223 (2001).}

\bibitem{ZG08}
 S. Zohren and R. Gill
 \href{http://dx.doi.org/10.1103/PhysRevLett.100.120406}{Phys. Rev. Lett. \textbf{100}, 120406 (2008).}

\bibitem{CGLMP02}
 D. Collins, N. Gisin, N. Linden, S. Massar, and S. Popescu,
 \href{http://prl.aps.org/abstract/PRL/v88/i4/e040404}{Phys. Rev. Lett. \textbf{88}, 040404 (2002).}

\bibitem{Masanes03}
 L. Masanes,
 Quant. Inf. Comp. \textbf{3}, 345 (2003).

\end{thebibliography}
\end{document}